\documentclass[twocolumn,showpacs,preprintnumbers,amsmath,amssymb]{revtex4}
\usepackage{amsmath,amssymb,graphics,epsfig,subfigure}
\usepackage{color}

\begin{document}

\thispagestyle{empty}

\begin{center}

\title{Thermodynamic Nature of Black Holes in Coexistence Region}

\date{\today}
\author{Shao-Wen Wei \footnote{E-mail: weishw@lzu.edu.cn},
Yu-Xiao Liu \footnote{E-mail: liuyx@lzu.edu.cn}}

\affiliation{$^{1}$Lanzhou Center for Theoretical Physics, Key Laboratory of Theoretical Physics of Gansu Province, and Key Laboratory of Quantum Theory and Applications of MoE, Lanzhou University, Lanzhou, Gansu 730000, China,\\
 $^{2}$Institute of Theoretical Physics $\&$ Research Center of Gravitation, Lanzhou University, Lanzhou 730000, People's Republic of China}

\begin{abstract}
Studying the system state of coexistence regions will peek into to reveal microscopic interactions between different phases of a thermodynamic system. However, there is no effective method to study thermodynamic nature of the coexistence black hole regions for the failure of the equation of state. Aiming at these coexistence states, in this work, we develop a general approach by introducing two new ratio parameters. The first one is the ratio of the horizon radii of the saturated coexistence small and large black holes, and the second one measures that of the small black hole molecule number to the total molecule number. We demonstrate that the first parameter can serve as an order parameter to characterize the first-order phase transition. The study also shows that the black hole state in the coexistence region is uniquely determined by these two introduced parameters bounded between 0 and 1. These results are quite significant in the analytical study of phase transition and the microscopic nature of black hole in the coexistence regions.
\end{abstract}

\keywords{Classical black hole, phase transition, equation of state, coexistence curve}

\pacs{04.70.Dy, 04.70.Bw, 05.70.Ce}

\maketitle
\end{center}

\section{Introduction}

Thermodynamics has been one of the active areas in modern black hole physics. The study of black hole thermodynamics shall provide insight into the quantum gravity and nature of black hole. Recently, it was extensively observed that distinct phase transitions exist in anti-de Sitter (AdS) black hole systems, where the cosmological constant was treated as pressure \cite{Kastor}. The small-large black hole phase transition \cite{Chamblin,Mann,Gunasekaran}, analogous to the gas-liquid phase transition of the van der Waals (VdW) fluid, was found to be universal in the charged AdS black hole systems. Further combining with the thermodynamical geometry, it was shown that both the repulsive and attractive interactions can dominate the two neighboring black hole molecules \cite{Weiw,Weilm,WeiWeiWei}. Other interesting issues including the Euler relation and dual conformal field theory have also been examined in Refs. \cite{Frassino,Ahmed,Gong,Kubiznak,Gao,XuWu,LiuLiu,BaiBai,MaPang}.

The starting point of black hole phase transition comes from the analogy of the equation of state (EoS) of the VdW fluid. For a general AdS black hole system, the EoS can be expressed as
\begin{eqnarray}
 P=\frac{T}{v}+\frac{f_{2}(\alpha; T)}{v^2}+\sum_{i\ge3}\frac{f_{i}(\alpha; T)}{v^i},\label{eos}
\end{eqnarray}
where $v$ is the specific volume of the system, and $f_{i}$ denote the functions that depend on the black hole parameters $\alpha$ and temperature. If taking $f_{2}=-a$, $f_{i\geq3}=0$ and replacing $v$ with $v-b$ in the first term on the right side of the equation, it shall recover the EoS of the VdW fluid model with $a$ and $b$ measuring the attraction and nonzero size of the molecules of the VdW fluid. Although the VdW fluid is developed to describe the phase transition, it has limitation in exploring the critical phenomena of an actual system. Significantly, since black holes have the similar EoS of the VdW fluid, one can naturally demonstrate that black hole phase transitions are of the VdW-like type. In fact, many works has confirmed this point. Besides, more interesting black hole phase diagrams are exposed. Nevertheless, black hole systems also hold their unique features. For examples, the gravitational constant $G$ appears in the functions $f_{i}$, which indicates that the EoS describes a gravity system. Adopting the Planck constant $l_{P}=1$, it is easy to find that the specific volume shown in (\ref{eos}) is $v\sim r_{h}$ with $r_{h}$ denoting the length scale of the black hole. This is quite different from the ordinary system, where the specific volume is proportional to the cubic power of the system characteristic length. Such difference leads to the inconsistence in the Maxwell equal area law in determining the phase transition point. In particular, the higher orders $f_{i\geq3}$ in the EoS for the black hole systems is quite universal. For the static and spherically symmetric black holes, entropy is proportional to the power of specific volume leading to the disappearance of the specific heat at constant volume. All these features reveal the quantum and gravity nature of black hole thermodynamics included in the EoS (\ref{eos}).

Despite the success, there is a huge challenge that the thermodynamical nature of the coexistence black hole region remains unknown, mainly caused by the failure of the EoS. Since in the coexistence region, the phenomena of the system are entirely determined by the coexistence and competition of the microscopic molecules. Therefore, studying it can help us insight into the unique microstructure of black holes. With this in mind, our goal is to investigate the universal properties of the small and large black holes in the coexistence region, which will lead to a comprehensive understanding of black hole thermodynamics within the phase space.

\begin{figure}
\includegraphics[width=6cm]{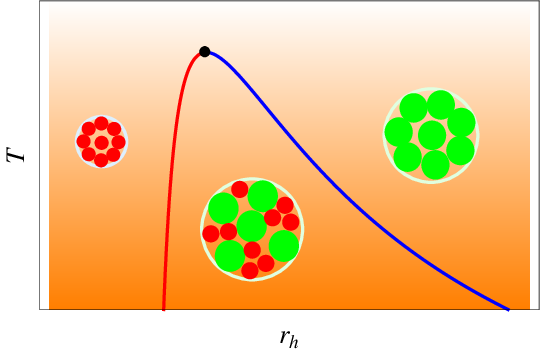}
\caption{Sketch of the phase diagram of the first-order phase transition near the critical point. The red and blue curves represent the statured coexistence small and large black holes, respectively. Black dot denotes the critical point. The small and large black holes locate at the left and right sides. Below the solid curve is the coexistence region where the small and large black holes coexist. The small circles represent different phase molecules.}\label{pPhaseDiag}
\end{figure}

As well known, the small and large black holes are characterized by the values of their horizon radii $r_{h}$. For clarity, we sketch the features of the small-large black hole phase diagram in the $T-r_h$ plane in Fig. \ref{pPhaseDiag}. The small and large black hole phases are, respectively, located at the left and right sides. Comparing with the large black hole, the small black hole admits small size molecules while high density \cite{Weiw}. Below the coexistence curve, constituted by the left red curve (coexistence saturated small black hole) and right blue curve (coexistence saturated large black hole), is the coexistence black hole region. As the EoS remains unknown in this particular region, previous studies have neglected the corresponding thermodynamic properties. However, we will demonstrate that these properties can be effectively explored by utilizing the properties of the coexistence saturated small and large black holes.

First, let us introduce a key ratio parameter
\begin{eqnarray}
 \epsilon=1-\frac{r_{hs}}{r_{hl}},\label{ratio}
\end{eqnarray}
where $r_{hs}$ and $r_{hl}$ denote the horizon radii of the coexistence saturated small and large black holes at the same temperature and pressure, respectively. They coincide and give $\epsilon=0$ at the critical point. Since $r_{hs}\leq r_{hl}$, one always has $\epsilon\in$ [0, 1]. More importantly, there are extremal black holes with vanished temperature. For this case, $ r_{hl}$ tends to infinity, while $ r_{hs}$ remains a finite value. As a result, we have $\epsilon=0$ for the extremal black holes. Further adjusting the parameters such that the black hole horizon disappears and a naked singularity is exposed, the thermodynamics will fail, so no any phase transition exists. We shall show that this ratio provides us with a favorable parameter for our following study. Here, we emphasize its advantages: i) It acts as an order parameter for characterizing the small-large black hole transitions. ii) The coexistence curves of the black hole and the VdW fluid can be parameterized analytically by $\epsilon$. iii) By further combining with the ratio of the molecule numbers of the small and large black holes, denoted by $x$, the state of the black hole system within the coexistence region can be uniquely determined. The effective EoS will also be given. On the other hand, the difference of the radii $\Delta=r_{hl}-r_{hs}$ is also an order parameter and can be used to characterize the small-large black hole phase transition. However, its value is unbounded and thus it is very difficult to parameterize coexistence curves via it.

The present work is organized as follows. In Sec. \ref{op}, we examine the ratio $\epsilon$ near the critical point and confirm that it indeed can act the order parameter to describe the small-large black hole phase transition. The analytical parameterized coexistence curve of the small and large black hole phase transitions, as well as the liquid and gas phase transition shall be given in Sec. \ref{apcc}. Then the equation of state in the coexistence region will be effectively investigated in Sec. \ref{spicr}. Finally, we summarize and discuss our results in Sec. \ref{con}.

\section{Order parameter}
\label{op}

It is extensively known that the difference of the thermodynamical volumes of the coexistence saturated small and large black holes is an order parameter for this VdW like phase transition. Here, we demonstrate that the ratio $\epsilon$ also acts as a significant order parameter.

Near the critical point, the reduced pressure can be expanded as \cite{Mann}
\begin{eqnarray}
 p=1+a_{10}t+a_{11}t\omega+a_{03}\omega^3+\mathcal{O}(t\omega^2, \omega^4),
\end{eqnarray}
where the coefficients $a_{10}$, $a_{11}$, and $a_{03}$ depend on the parameters of the specific black holes. The reduced quantities are defined as $p=P/P_{c}$, $t=T/T_{c}-1$, and $\omega=V/V_{c}-1$. Solving the Maxwell equal area law $\oint\omega dp=0$, or equivalently $\int_{\omega_{s}}^{\omega_{l}}(\omega\partial_{\omega}p) d\omega=0$, we have
\begin{eqnarray}
 a_{11}t \omega_{l}^2+\frac{3}{2}a_{03}\omega_{l}^4=a_{11}t \omega_{s}^2+\frac{3}{2}a_{03}\omega_{s}^4.
\end{eqnarray}
Along the coexistence curve, the EoS holds for the saturated small and large black holes
\begin{eqnarray}
 p=1+a_{10}t+a_{11}t\omega_{s}+a_{03}\omega_{s}^3,\\
 p=1+a_{10}t+a_{11}t\omega_{l}+a_{03}\omega_{l}^3.
\end{eqnarray}
Combining with these equations, it is easy to obtain
\begin{eqnarray}
 \omega_{s,l}=\mp\sqrt{\frac{a_{11}}{a_{03}}}\sqrt{-t},\label{los}
\end{eqnarray}
for the coexistence saturated small and large black holes.

From (\ref{los}), one arrives
\begin{eqnarray}
 \epsilon&=&1-\left(\frac{\omega_{s}+1}{\omega_{l}+1}\right)^{\frac{1}{d-3}}\nonumber\\
 &=&1-\left(\frac{a_{03}-a_{11}t-2\sqrt{-a_{03}a_{11}t}}{a_{03}+a_{11}t}\right)^{\frac{1}{d-1}},
\end{eqnarray}
for $d$-dimensional case. Note that we have used
\begin{eqnarray}
 \frac{r_{hs}}{r_{hl}}=\left(\frac{V_{s}}{V_{l}}\right)^{\frac{1}{d-1}}.
\end{eqnarray}
Expanding it near the critical point $t=0$, we obtain
\begin{eqnarray}
 \epsilon=\frac{2}{d-1}\sqrt{\frac{a_{11}}{a_{03}}}\sqrt{-t}+\mathcal{O}(t).\label{ess}
\end{eqnarray}
Obviously, $\epsilon$ exhibits a critical exponent 1/2, strongly indicating that it can serve an order parameter, and can be utilized to characterize the small-large black hole phase transition. This shall be further confirmed by the results of the VdW fluid and charged AdS black holes in what follows.

Before ending this section, we would like to give a note on the Maxwell equal area law. Actually, it mainly lies in the equality of the free energy of the coexistence saturated small and large black holes. Meanwhile, one should note that $\omega$ is related with the thermodynamical volume rather than the specific volume. The details can be found in Ref. \cite{Wei3}. For the VdW fluid system, one can obtain its free energy by combining with the micro-model. However, for the black hole systems, since the micro-model is unclear, one could not derive its free energy via the EoS or the Hawking temperature. For a specific black hole system, its free energy can be calculated through the action.

\section{Analytical parameterized coexistence curve}
\label{apcc}

Now, we turn to analytically study the coexistence curve by using the parameter $\epsilon$. Let us focus on the charged AdS black hole. In Ref. \cite{Chamblin}, it was found that the free energy demonstrates the swallow tail behaviors indicating the existence of the phase transition. Treating the cosmological constant as the pressure, the small-large black hole phase transition of the VdW type was observed in the extended phase transition \cite{Mann}. Starting with the $d$-dimensional case, the EoS in the reduced parameter space was given by \cite{Gunasekaran}
\begin{eqnarray}
 p=\frac{4(d-2)\tau}{(2d-5)\tilde{r}_{h}}-\frac{d-2}{(d-3)\tilde{r}_{h}^2}+\frac{1}{(d-3)(2d-5)\tilde{r}_{h}^{2d-4}}.
\end{eqnarray}
Similar to the VdW fluid, the black hole charge has been scaled out in the reduced parameter space. It is worthwhile noting that, the Maxwell equal area law is in the form of $\oint\tilde{r}_{h}^{(d-1)} dp=0$ for the $d$-dimensional charged AdS black hole \cite{Wei3}. By making use of it, the analytical parameterized curve can be obtained for any $d$. However, its expression is long and so we will not show it here \cite{kkk}. For clarity, we list the result for $d=4$
\begin{eqnarray}
 p&=&\frac{36 (1-\epsilon)^2}{(\epsilon^2 -6 \epsilon +6)^2},\label{pad4}\\
 \tau&=&\frac{3 \sqrt{6} (2-\epsilon) (1-\epsilon)}{(\epsilon^2-6\epsilon
   +6)^{3/2}}.\label{tad4}
\end{eqnarray}
and $d$=5
\begin{eqnarray}
 p&=&\frac{3 \sqrt{15} (1-\epsilon)^2 (\epsilon^2 -5  \epsilon
   +5)}{(\epsilon ^4-7 \epsilon ^3+22 \epsilon ^2-30 \epsilon +15)^{3/2}},\\
 \tau&=&\frac{15 \sqrt[4]{15} (2-\epsilon)^3 (1-\epsilon)}{8 (\epsilon ^4-7 \epsilon ^3+22 \epsilon ^2-30 \epsilon +15)^{5/4}}.
\end{eqnarray}
Fortunately, for the four-dimensional case, we can solve the ratio $\epsilon$ from the pressure (\ref{pad4}), and then express the temperature as
\begin{eqnarray}
 \tau=\sqrt{\frac{1}{2}p(3-\sqrt{p})},
\end{eqnarray}
which is exactly the result given in Ref. \cite{Spallucci}. Adopting the analytical parameterized formulas, one can easily obtain the corresponding phase diagram for the black holes. Moreover, we find that when the dimension number $d$ is larger than 100, the phase diagrams almost hold unchanged in the reduced parameter space.

Employing with the parameterized formula, we can expand the ratio near the critical temperature as
\begin{eqnarray}
 \epsilon=\frac{2\sqrt{6}}{\sqrt{2d-5}}(1-\tau)^{\frac{1}{2}}-\frac{12}{2d-5}(1-\tau)+\mathcal{O}((1-\tau)^{\frac{3}{2}}),\label{spis}
\end{eqnarray}
which exactly confirms the result (\ref{ess}), and indicates that the ratio $\epsilon$ acts as an order parameter for any dimension $d$.

\begin{figure}
\includegraphics[width=6.5cm]{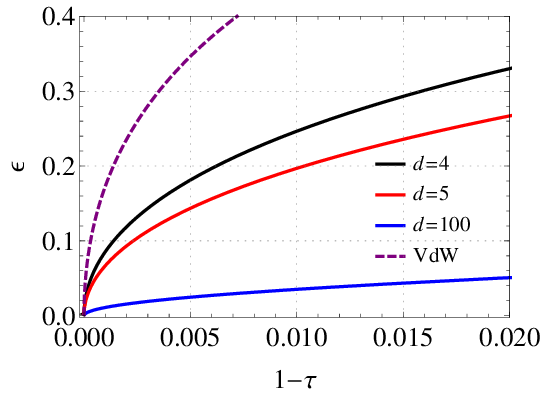}
\caption{The ratio $\epsilon$ as a function of the shifted temperature 1-$\tau$. The critical behavior can be observed near the coordinate origin.}\label{ptauepsion}
\end{figure}

More interestingly, applying the study to the VdW fluid, we obtain the analytical parameterized coexistence curve for the VdW fluid
\begin{eqnarray}
 p &=&
 \frac{27 (1-\epsilon) \left[2 \epsilon+(2-\epsilon) \ell_\epsilon \right]^2
   \left[\epsilon ^2-(1-\epsilon) \ell_\epsilon^2\right]}{\epsilon
   ^2 \left[\epsilon +\ell_\epsilon\right]^2 \left[\epsilon+(1+\epsilon) \ell_\epsilon\right]^2},\\
  \tau &=&
 \frac{27 \left[(\epsilon -2) \ell_\epsilon-2 \epsilon \right]
    \left[(2-\epsilon)  \epsilon +2 (1-\epsilon) \ell_\epsilon\right]^2}
   {8 \epsilon
   \left[\epsilon +\ell_\epsilon\right]^2
   \left[\epsilon -(\epsilon -1) \ell_\epsilon\right]^2},\label{ww}
\end{eqnarray}
where $\ell_\epsilon=\ln(1-\epsilon )$.
When $\epsilon=1-\frac{\nu_{l}-\frac{1}{3}}{\nu_{g}-\frac{1}{3}}\rightarrow1$, it gives $p=\tau=0$, and when $\epsilon\rightarrow0$, the critical point at $p=\tau=1$ is given. Employing this analytical formula, it is possible to precisely study various properties of the VdW fluid in all parameter spaces without using any approximation. This approach shall represent a significant improvement over earlier approximate studies \cite{Lekner,Berberan}, which are limited to the $\tau \rightarrow$ 0 and 1 regimes.

Interestingly, the ratio parameter $\epsilon$ can also be obtained by expanding near the critical temperature
\begin{eqnarray}
 \epsilon=6\sqrt{1-\tau}-18(1-\tau)+\mathcal{O}((1-\tau)^{\frac{3}{2}}).
\end{eqnarray}

Moreover, we also show the ratio $\epsilon$ as a function of $1-\tau$ in Fig. \ref{ptauepsion}. It is evident that, for the VdW fluid and charged AdS black holes with dimensions $d$ = 4 and 5, $\epsilon$ approaches zero as $(1-\tau)^{\frac{1}{2}}$.

To summarize, the ratio $\epsilon$ that we have introduced can serve as an order parameter to characterize the small-large black hole phase transition. Significantly, the coexistence curves for both the VdW fluid and $d$-dimensional charged AdS black holes can be analytically obtained in parameterized forms. In the following sections, we shall demonstrate that this ratio can also serve as a parameter for characterizing the black hole state within the coexistence region.

\section{State parameters in coexistence region}
\label{spicr}

The exploration of the coexistence region in black hole systems has been neglected in previous studies due to the violation of the EoS. Now, we aim to explore this issue by utilizing the ratio $\epsilon$ and taking the charged AdS black hole as an illustrative example.

\begin{figure}
\includegraphics[width=6.5cm]{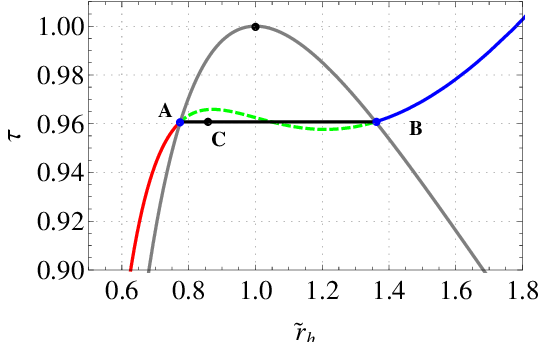}
\caption{The isobaric curve with $p$=0.9 for the four-dimensional charged AdS black hole. The gray solid curve is the coexistence curve and the black dot is the critical point. Points A and B denote two saturated coexistence small and large black holes. Point C in the coexistence region represents a state of coexistence small and large black holes sharing with the same temperature and pressure as points A and B. The phase transition from saturated small black hole to large black hole is described by the horizontal line.}\label{pTs}
\end{figure}

Let us consider one isobaric curve with $p=0.9$ shown for $d$=4 dimensional charged AdS black hole in Fig. \ref{pTs}. The gray curve denotes the coexistence curve, and below which is the coexistence region. The oscillatory part of the isobaric curve in green color should be replaced by a horizontal line, and which describes a phase transition between the small and large black holes. Points A and B are the saturated small and large black hole states. Meanwhile, each point on the horizontal line represents a system state with the same temperature and pressure. For example, the state denoted by point C shares the same temperature and pressure of the states A and B, which are not independent and given in (\ref{pad4}) and (\ref{tad4}), respectively. The ratio $\epsilon$ serves as one characteristic parameter of state C, but only using it cannot uniquely determine the state. Since state C actually contains both the coexistence small and large black holes. Hence, a second parameter is needed to characterize this property.

During the transition from state A to B, the molecules of the saturated small black hole gradually transform into those of the saturated large black hole. To characterize this property, we introduce another parameter $x=N_{s}/(N_{s}+N_{l})$, which measures the number of the saturated small black hole molecules to the total number of molecules. Further combining with the lever rule
\begin{eqnarray}
 \frac{N_{s}}{N_{l}}=\frac{\tilde{V}_{hl}-\tilde{V}_{hi}}{\tilde{V}_{hi}-\tilde{V}_{hs}},
\end{eqnarray}
we can express the parameter $\tilde{r}_{hi}$ of an intermediate state, such as state C, as
\begin{eqnarray}
 \tilde{r}_{hi}=\frac{\sqrt{\epsilon^2 -6\epsilon +6}}{\sqrt{6}
   (1-\epsilon)}\sqrt[3]{1-x(1-(1-\epsilon)^{3})}.
\end{eqnarray}
Here $\tilde{r}_{hi}$ is bounded by the radii of the saturated small and large black holes. We need to point out that it just denotes the location of the system in the phase diagram instead of the actual radius of the coexistence black holes. It is easy to find that $x=$1 and 0 correspond to states A and B, respectively. So, analogous to the fluid systems, $x$ gives the ratio of the saturated small and large black hole molecules located at this system state. On the other hand, understanding these coexistence black holes from a gravitational perspective is worth future investigation via the potential relationship between the black hole thermodynamics and gravity.

\begin{figure}
\includegraphics[width=6.5cm]{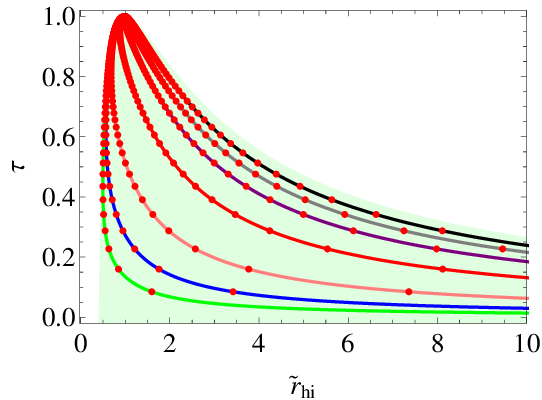}
\caption{Coexistence states characterized by the ratios $\epsilon$ and $x$ shown in $\tau$-$\tilde{r}_{hi}$. These curves are plotted by taking $x$=0.3, 0.5, 0.7, 0.9, 0.99, 0.999, 0.9999 from right to left. They start at the critical point with $\epsilon$=0 and extend to the right with the increase of $\epsilon$. The coexistence region is marked in light green color.}\label{pTaur}
\end{figure}

Finally, we find that each coexistence state can be parameterized by two quantities $\epsilon$ and $x$, both of which are bounded between 0 and 1. In Fig. \ref{pTaur}, we show these coexistence states by taking $x$=0.3, 0.5, 0.7, 0.9, 0.99, 0.999, 0.9999 from right to left. The coexistence region is marked in light green color. Note that the parameter $x$ heavily relies on the saturated small and large black holes and is not directly related to the extremal black holes or naked singularities.

All of these curves coincide at the critical temperature with $\epsilon$=0, and extend to the right. In particular, with the increase of $x$, the curve shifts towards the saturated small black hole curve. The monotonic behavior is also broken. Nevertheless, each state within the coexistence region is characterized by a pair of parameters, namely ($\epsilon$, $x$). This study can also be easily generalized to higher dimensional cases.

\section{Conclusions}
\label{con}

In this study, we have investigated the thermodynamic nature of the coexisting black hole phases of the VdW type by introducing two new ratio parameters, $\epsilon$ and $x$. Each system state within the coexistence region can be uniquely determined by these parameters, offering an opportunity to explore the coexisting system states that have been previously overlooked.

Our findings suggest that the ratio $\epsilon$ plays a crucial role in the study of coexisting physics. By utilizing this parameter, we obtained, for the first time, an analytical parameterized form for the coexisting saturated liquid and gas phases. This shall greatly improve the fitting formula given by Ref. \cite{Johnston}. Consequently, all other properties of the VdW fluid can be analyzed analytically without any approximation. For the $d$-dimensional charged AdS black holes, the analytical parameterized form of the coexistence curve was also obtained. By performing the general calculations, we demonstrated that the ratio $\epsilon$ can serve as an order parameter for characterizing the liquid-gas phase transition or the small-large black hole phase transition. This conclusion is further supported by the results of the VdW fluid and charged AdS black holes.

Employing with the lever rule, we introduced another parameter, $x$, which measures the ratio of the small black hole molecule number to the total molecule number. Our result indicates that a black hole state within the coexistence region can be uniquely determined by the ratios $\epsilon$ and $x$, thereby providing an effective EoS. This approach offers insight into the coexistent physics that has been neglected in previous studies.

While our study is limited to the VdW type phase transition, the approach we have developed can be generalized to the local analysis of other phase structures near the critical point, representing a second-order phase transition. Actually, this feature is quite prevalent in black hole chemistry.

On the other hand, although we gave an effective EoS in the coexistence region, several issues are still not resolved. For examples, the first law still remains to be established and the free energy corresponding to the effective EoS is still unclear. We expect to address these issues in future study.

In conclusion, we have developed a general analytical approach to gain insight into the coexistent physics. The effective EoS represented by the two ratio parameters can be used to test other underlying properties.

\section{Acknowledgements}
This work was supported by the National Natural Science Foundation of China (Grants No. 12075103, No. 11875151, and No. 12247101) and the Major Science and Technology Projects of Gansu Province.


\begin{thebibliography}{99}
\bibitem{kkk}
 By treating the ratio of the radii of the coexisting small and large black holes as an intermediate variable, an equivalent coexistence curve was presented in Ref. \cite{Zhang}.
\end{thebibliography}
\end{document}